# An environmental dose experiment

Luis Peralta

Faculdade de Ciências da Universidade de Lisboa and Laboratório de Instrumentação e Física Experimental de Partículas, Lisboa, Portugal

**Abstract**

Worldwide several radiation sources contribute to the delivered dose to the human population. This radiation also acts as natural background when detecting radiation, for instance from radioactive sources. In this work a medium-size plastic scintillation detector is used to evaluate the dose delivered by natural radiation sources. Calibration of the detector involved the use of radioactive sources and Monte Carlo simulation of the energy deposition per disintegration. A measurement of the annual dose due to background radiation to the body was then estimated. A dose value compatible with the value reported by the United Nations Scientific Committee on the Effects of Atomic Radiation was obtained.

**Introduction**

Living beings are exposed to natural radiation sources. The main natural sources of exposure are cosmic radiation and natural radionuclide found in the soil and in rocks. At ground level, muons (with energies in the range 1 to 20 GeV) constitute the dominant component of the cosmic ray field. Gamma photons are on the other hand the major contributor to dose due to radionuclides. Effective dose may vary from place to place but usually are in the 1 to 10 mSv/a [1], with an average at 2.4 mSv/a. Accordingly to UNSCEAR 2008 report [2], low-ionizing density radiation is responsible for an estimated 0.48 mSv/a from external terrestrial radiation to the body and 0.39 mSv/a from cosmic radiation (world average). We should note that these estimates have changed throughout the years and different sources quote different values.

The measurement of the environmental dose due to cosmic ray and gamma ray sources can be made using a plastic scintillation detector. In this work we describe the operation of a plastic scintillation detector used as an environmental dosimeter, assessing the annual dose delivered in the laboratory. The detector was built in our laboratory by students as part of a discipline project.

**Measuring radiation dose**

The environmental dose is due to weak sources and measurements are usually made by detectors with sufficiently large volumes. A popular choice is plastic scintillation detectors. Comparing to scintillation inorganic crystals, plastic scintillators have much lower price per unit volume. On the physical side, plastics have much better equivalence to water than inorganic crystals. When translating the dose $D_m$ delivered by photons at the dosimeter material to dose $D_w$ at water (or organic tissue) the conversion factor is the ratio of average mass absorption coefficients $\overline{\mu_{en}/\rho}$ [3,4]

$$D_w = D_m \overline{(\mu_{en}/\rho)_m}/\overline{(\mu_{en}/\rho)_w}. \qquad (1)$$

The average value must be obtained over the relevant photon fluence spectrum [5], the details of which may be largely unknown for natural sources. If the ratio of mass absorption coefficients is approximately constant over that range, the knowledge on the spectrum is no longer relevant. This is the case of some plastics like PMMA, Polystyrene or PVT as it can be checked using data from NIST website [6]. In table 1 the ratio of mass absorption coefficients of these plastics to water is presented.

Table 1. Approximate ratio of mass absorption coefficients of plastics to water in the 0.3 to 4 MeV range. The ratio for 1.25 MeV is presented. Mass absorption coefficients were obtained from [6].

|  | PMMA | Polystyrene | PVT |
|---|---|---|---|
| $(\mu_{en}/\rho)_m/(\mu_{en}/\rho)_w$ | 0.9376 | 0.9693 | 0.9761 |

For charged particles equation 1 doesn't apply and the relevant conversion factor is given by the mass stopping power $\overline{(S/\rho)}$ coefficients ratio [3,4]

$$D_w = D_m \overline{(S/\rho)_m}/\overline{(S/\rho)_w}. \qquad (2)$$

The main component of cosmic rays reaching the Earth's surface is due to high-energy muons with energy in excess of 1 GeV [7]. These particles are often refereed as minimum ionizing particles (mip), since their stopping power is approximately equal to the minimum value obtained from the Bethe-Bloch curve [7]. The ratio of mass stopping power coefficients of plastics to water for minimum ionizing particles is presented in table 2.

Table 2. Ratio of mass stopping power coefficients of plastics to water for minimum ionizing particles [7].

|  | PMMA | Polystyrene | PVT |
|---|---|---|---|
| $(S/\rho)_m/(S/\rho)_w$ | 0.9684 | 0.9719 | 0.9819 |

**The detector**

Our detector's sensitive volume is a cylinder, 4.60 cm in diameter and 10.00 cm long made of PMMA scintillator (RP-200A [8]). The scintillator is coupled with optical grease to a 2 inch PMT (Hamamatsu R6231). The scintillator is wrapped in Tyvek (trade mark) white paper for better light collection. The ensemble was enclosed in a PVC case with 8 mm-thick walls. The signal from the detector was fed to a NIM amplifier and collected by a multichannel analyser. To calibrate the detector radioactive sources ($^{137}$Cs, $^{60}$Co and $^{22}$Na) were used. The sources' activity were measured in our laboratory using a 3"×3" NaI(Tl) detector following the procedure in [9]. The measured values are presented in table 3.

Table 3. Measured activities of the used radioactive sources. Uncertainties on the activity are 5%. Half-lives obtained from [10].

| Source | Half-life (year) | Activity (kBq) |
|---|---|---|
| $^{137}$Cs | 30.07 | 133 |
| $^{60}$Co | 5.2714 | 339 |
| $^{22}$Na | 2.6019 | 1783 |

To obtain the deposited energy in the detector per disintegration a Monte Carlo simulation of the set-up was made using Penelope [11-13]. Penelope simulates the transport of electrons, photons and positrons through matter in a wide energy range (250 eV to 1 GeV). The cylindrical geometry of the detector enables the use of *pencyl* routine with a less complicated geometry interface than standard *penmain* routine [11]. A full description of the detector (scintillator and detector wall) was made. Point-like radioactive sources, placed at chosen distances were simulated (one distance at a time). For each radioactive source, the relevant photon spectrum was considered (see table 4). The program is designed to generate and track one primary particle per event, while on average more than one photon can be emitted per event in most cases. One way to correctly account for all deposited energy is to separately simulate each photon and add-up each contribution weighted by the photon's percentage per disintegration. While one should consider isotropic sources and thus 4π photon emission, for the two 511 keV annihilation photons, only one photon should be simulated in a 2π emission and thus increasing its weight by a factor 2. The program output will be the deposited energy in the detector sensitive volume per event (or per disintegration). The dose is computed as the ratio of the deposited energy to the scintillator mass. Considering a density of 1.19 g cm$^{-3}$ for PMMA [8] the mass of our scintillator is (0.198 ±0.001) kg. The experiment final result will critically depend on the accuracy of the Monte Carlo simulation. In most cases the best way to assess the quality of the simulation is to compare simulated results with experiment. An experiment using a low-activity, large solid angle $^{40}$K source, was designed for this purpose. The comparison between simulated and measured dose is made, providing a cross check for the method.

Table 4. Photon energy and percentage per disintegration for $^{137}$Cs, $^{60}$Co and $^{22}$Na. Data adapted from [10] except for 511 keV photon from $^{22}$Na [14].

| Source | Photon energy (keV) | Percentage per disintegration |
| --- | --- | --- |
| $^{137}$Cs | 32.05 | 5.8 |
|  | 36.51 | 1.25 |
|  | 661.65 | 85.1 |
| $^{60}$Co | 1173.24 | 99.97 |
|  | 1332.50 | 99.99 |
| $^{22}$Na | 511. | 90.35 |
|  | 1274.53 | 99.94 |

**Acquisition and data processing**

Plastic scintillation detectors have poor energy resolution [15], as demonstrated in the spectrum of figure 1 obtained with our detector for a $^{137}$Cs source. The gamma background spectrum presents a similar shape (i.e. no peaks are present) although extended to higher energy (see figure 3). Both low density and moderate light yield (when compared to inorganic scintillation crystals) contribute to this outcome. However for dosimetry this is not a serious disadvantage, since the measurement doesn't rely on single photon detection but rather on the integration of the energy over a number of events. Thus, it is important to maintain signal-energy proportionality, but not on an event basis.

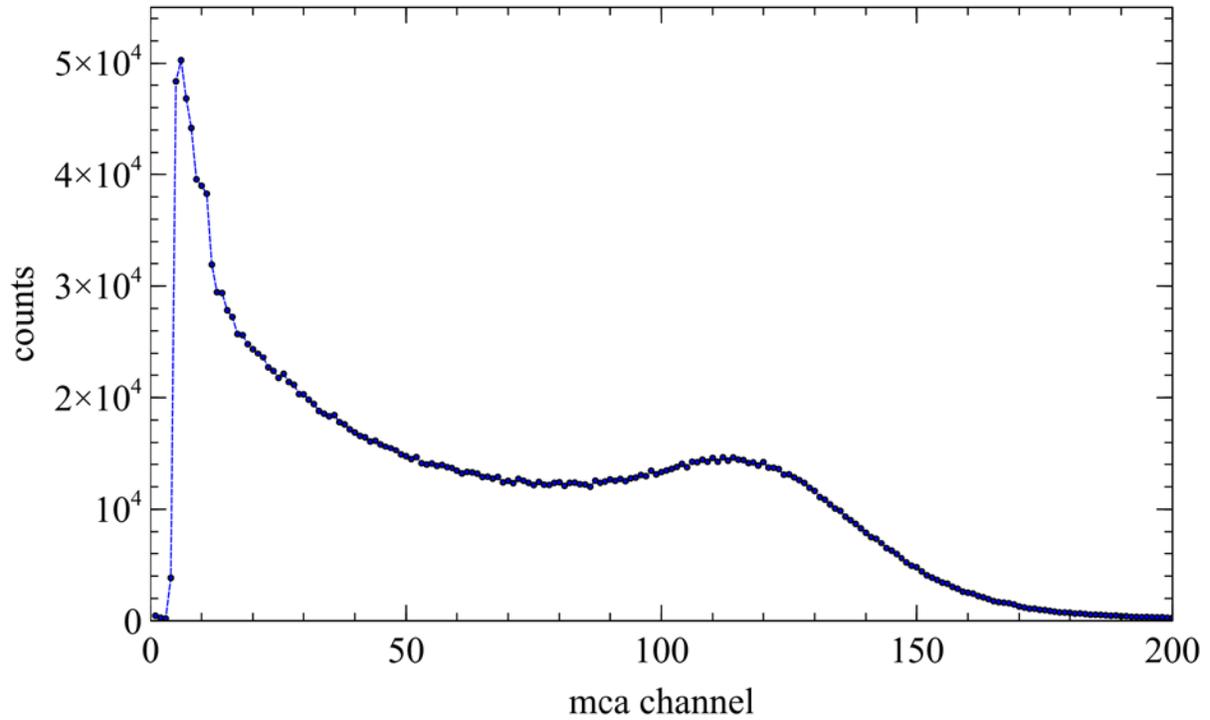

Figure 1: $^{137}$Cs spectrum obtained with the plastic scintillation detector.

In this experiment the signal *S* is obtained as the sum of product of the multichannel analyser (mca) channel *ch* (proportional to single event energy) by the number of counts in that channel, in a given channel range, average over acquisition time T, ie.

$$S = \frac{1}{T}\sum_{ch_{min}}^{ch_{max}}(ch \times counts). \tag{3}$$

The choice of the lower threshold channel $ch_{min}$ depends on electronic background noise. The evaluation of this threshold is crucial for the environmental dose determination since electronic and natural backgrounds are mixed. Once the threshold is set, the calibration of the detector can be made. Figure 2 displays dose per disintegration versus signal for a set of measurements with each of the radioactive sources placed at different distances from the detector. A linear relation between the two variables is observed over several orders of magnitude, which is a remarkable result.

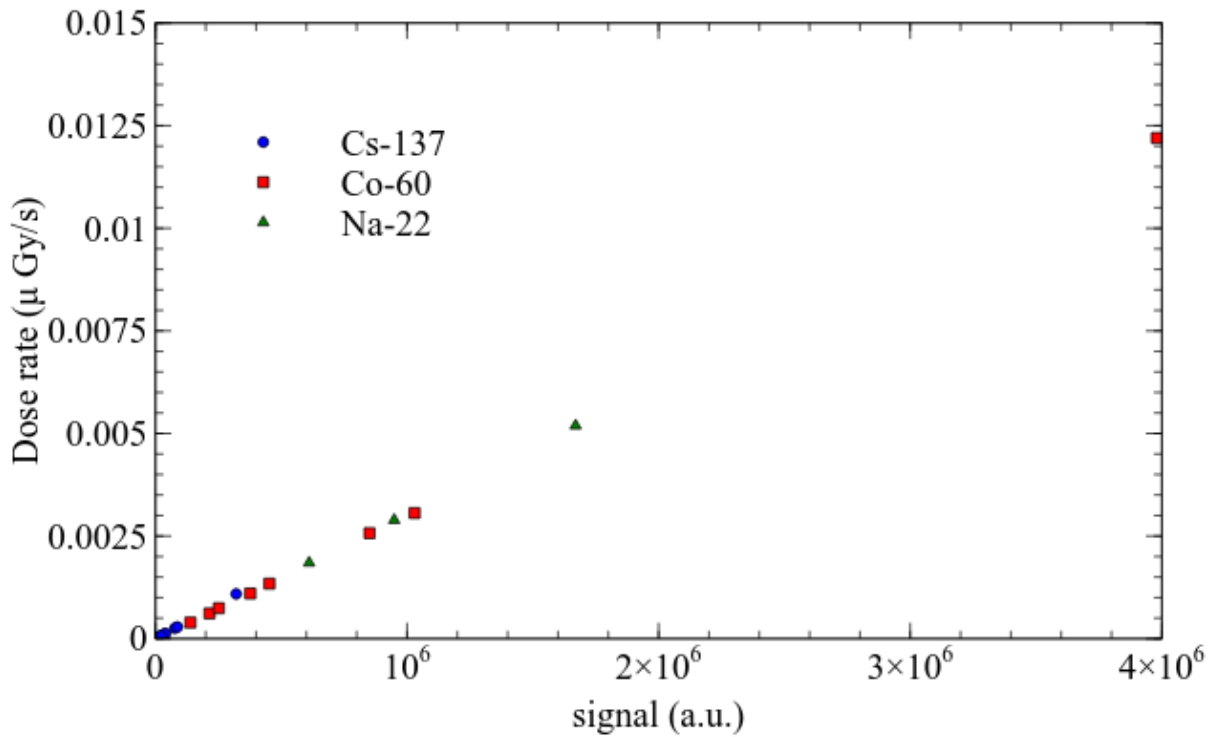

Figure 2: Calibration curve of the scintillation detector using three radioactive sources placed at different distances from the detector. A low threshold of 30 channels was applied to each spectrum, as explained in the text.

**Electronic background assessment**

It's a well established fact that PMTs have dark noise current [15], contributing to the low energy noise. After detector's assemblage it's not easy to separate this noise from the background originated by natural radiation sources. An assessment of its importance in percentage and range can be made comparing the background spectra (i.e. without the presence of individual sources) obtained with the detector inside and outside a lead housing. For the task, a lead brick (5 cm thick) castle was built in such a way that the detector could be fully enclosed. In figure 3 it is presented the result for the background spectra obtained with the detector outside and inside the lead castle. Analysing the ratio of the two spectra, two distinct regions can be established. A low energy region, with growing ratio values and a fairly constant ratio region above the threshold channel.

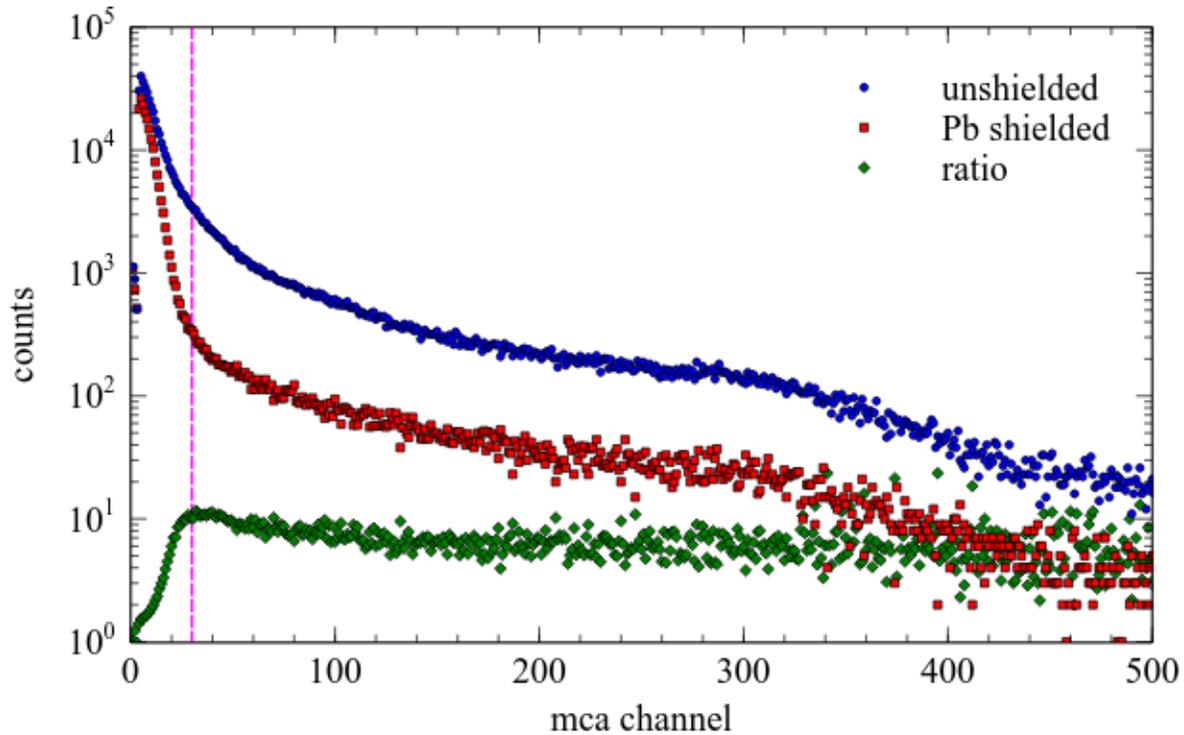

Figure 3: Background spectra obtained with the detector outside and inside the lead castle. The ratio of the spectra is also presented. A dashed line is plotted at the threshold channel.

The low energy gamma background is greatly reduced by the lead shield, so in the low channel range electronic noise contribution to the spectrum is enhanced. As energy increases so does the percentage of radiation penetrating the shielding. For high enough channel values the contribution of electronic noise to the background (which is not affected by the lead shielding) is greatly reduced and the ratio tends to be fairly constant. A threshold was then arbitrarily set at the channel with 90% of the maximum ratio value (corresponding to channel 30 in our case). A relative difference of 3% is observed in the overall ratio (i.e. the ratio of the spectra integrals) when the threshold is moved 5 channels (left or right).The threshold ensures a considerable reduction in the electronic noise contribution at the expense of cutting the contribution of low energy radiation to the measured dose value. The background signal eliminated by this cut was approximately 20% of the total measured background signal. This 20% includes electronic noise background as well as background signal due to low energy deposits in the detector from gammas and muons.

A simple straight line fit was applied to the data in figure 2. A slope of $(3.02 \pm 0.04) \times 10^{-9}$ µGy/s and an intercept of $(2.9 \pm 1.7) \times 10^{-6}$ µGy/s were obtained for our setup.

**Muon contribution**

Muons from the pion decay, produced in the collisions of primary cosmic ray particles have average energy of the order of 3 GeV at ground level [7,16] and can be considered as minimum ionizing particles (mip) [7,17]. Since stopping power $dE/dx$ is a slow varying function of the particle energy for mips, the energy loss ΔE in thin slabs can be approximate by

$$\Delta E = \left(-\frac{dE}{dx}\right)\Delta x, \tag{4}$$

where $\Delta x$ is the thickness of the slab. For this purpose a slab can be considered as "thin" for a particle of energy $E$ if $\Delta E \ll E$. The mass stopping power for mip for PMMA is 1.929 MeV cm$^2$g$^{-1}$ [18]. Since the density of PMMA is 1.19 g cm$^{-3}$, a muon traversing the full length of the detector (10 cm) will then lose the energy amount of $\Delta E = 1.929 \times 1.19 \times 10 \approx 23$ MeV. We can thus expect to have muon contribution to the spectrum up to high energy. Furthermore, it is known [15] that the amount of scintillation light produced per deposited MeV is different for electrons (or gammas) and heavier particles. The calibration made using gamma sources is thus not applicable to extract the muon contribution to the dose. To our knowledge, there is no published work on the scintillation light produced by muons on the RP-200A scintillator. A separation of the muon signal component from the gamma is thus necessary.

To do so, information on the percentage of each radiation component is needed. That information cannot be obtained directly from the detector. However, the detected gamma component can be decreased to a negligible level shielding the detector with lead bricks. A 5 cm lead brick will attenuate the gamma component by $e^{-5\mu}$, where $\mu$ is the linear attenuation coefficient in cm$^{-1}$. For $^{60}$Co gamma photons the attenuation factor is about 0.036, being smaller for lower energy photons.

On the other hand high-energy muons still reach the detector. In fact, using 1.122 MeV g$^{-1}$ cm$^2$ for minimum ionizing particles [7, 18] in lead, a value of approximately 64 MeV for the lost energy is obtained. Let then *m* and *g* be, respectively the muon and gamma total signal above threshold as defined by equation 3. Assuming zero gamma contribution in the shielded scenario, one gets for the ratio *r* between shielded and unshielded cases

$$r = \frac{m}{m+g} \qquad (5)$$

which is the muon fraction contribution for the signal. For our data $r = 0.161 \pm 0.001$ (statistical uncertainty only).

**Dose due to background radiation**

The 30 channel threshold was applied to the background radiation spectrum and the signal obtained accordingly to equation 3. Applying calibration to the gamma component (i.e.83.9% of the signal) a value of $(1.64 \pm 0.02)$ µGy/s is obtained, corresponding to $(0.52 \pm 0.05)$ mGy for the time period of one year. This is the dose in PMMA. To convert dose in PMMA to dose in water equation 1 should be applied. Using $(\mu_{en}/\rho)_{pmma}/(\mu_{en}/\rho)_{water} = 0.9376$ from table 1, we arrive to $(0.48 \pm 0.05)$ mGy for dose in water. Only statistical uncertainties have been considered.

**Dose measurement of a low activity source**

The detector was also tested on dose measurement of a known low activity source. For this purpose potassium chloride (KCl) in powder was chosen. $^{40}$K is a natural occurring radioactive isotope of potassium in an abundance of 0.0117% and has an half-life of $1.277 \times 10^9$ year [10]. Potassium chloride can be purchased with high purity and exempt of any license. A cylindrical source weighting 3.040 kg was made. The source has a hole in its centre to accommodate the detector and its height was such that it encompasses the whole sensitive part of the detector. To compute the source activity the following equation can be used [19],

$$A = \frac{\ln(2)}{T_{1/2}} \times \frac{M_{KCl}B}{A_{KCl}} \times N_A \qquad (6)$$

where $T_{1/2}$ is the $^{40}$K half-life, $M_{KCL}$ the potassium chloride mass, *B* the $^{40}$K abundance, $A_{KCl}$ the molar mass of potassium chloride and $N_A$ the Avogadro's constant. For our source an activity value of $(49.4 \pm 0.5)$ kBq was computed. The experimental setup (cylindrical KCl source and scintillation detector) was simulated with Penelope program to obtain the dose per disintegration factor. For this purpose a density of $(0.93 \pm 0.02)$ g cm$^{-3}$ was estimated considering the measured weight and volume occupied by the KCl sample. A factor of $(0.960 \pm 0.003)$ keV/decay was obtained from simulation. To get the factor that translates activity to dose rate we multiply by $1.602 \times 10^{-16}$ J keV$^{-1}$, converting keV to joule and divide by the detector's mass. A factor of $(7.77 \pm 0.03) \times 10^{-16}$ Gy/decay is then obtained. Using this factor and the computed source activity a dose rate of $(3.84 \pm 0.03) \times 10^{-11}$ Gy/s is expected. For the used KCl source, after background subtraction to the signal, computed according to equation 3, and using the previous signal to dose rate calibration, a corresponding dose rate of $(3.5 \pm 0.2) \times 10^{-11}$ Gy/s was obtained, representing a difference of 9% to the computed value.

**Conclusion**

Background radiation is a constant presence everywhere. For most laboratory experiments it is just one more difficulty to deal with. In this work a plastic scintillator detector was used to estimate the annual dose due to background radiation. The measured dose rate value of 0.48 mGy/a is in the same range as the background dose due to gamma radiation from natural sources reported on international reports West Europe [2]. The same method was applied to the measurement of the dose delivered by a low activity KCl cylindrical source. A difference of 9% between computed and measured value was found, being a reasonable result given the overall uncertainties.

**Acknowledgements**

We are grateful to Laboratório de Instrumentação e Física Experimental de Partículas for the financial support given to this work. We are also grateful to Ashley Peralta for the review of the English text and to Prof. Jorge Sampaio for the critical reading of the document.